\documentclass[prl,twocolumn,showpacs,floatfix]{revtex4}
\usepackage{graphicx,amsfonts,amssymb,amsmath, hyperref}

\newif\ifhyper
\hypertrue
\ifhyper
\hypersetup{
   citecolor = {green},
   colorlinks = {true}, 
   urlcolor = {blue} 
}
\fi

\newcommand{\beq}{\begin{equation}}
\newcommand{\eeq}{\end{equation}}
\newcommand{\beqa}{\begin{eqnarray}}
\newcommand{\eeqa}{\end{eqnarray}}
\newcommand{\ket} [1] {\vert #1 \rangle}
\newcommand{\bra} [1] {\langle #1 \vert}
\newcommand{\braket}[2]{\langle #1 | #2 \rangle}

\newcommand{\widebar}[1]{\overline{#1}}

\def\bra#1{\langle#1\vert}
\def\ket#1{\vert#1\rangle}

\def\Longarrow{\protect\@lra}
\def\@lra{\relbar\joinrel\relbar\joinrel\relbar\joinrel%
          \relbar\joinrel\rightarrow}

\begin{document}

\title{Entanglement and SU($n$) symmetry in one-dimensional valence bond solid states}

\author{Rom\'an Or\'us}
\author{Hong-Hao Tu}

\affiliation{Max-Planck-Institut f\"ur Quantenoptik, Hans-Kopfermann-Stra\ss e 1, 85748 Garching, Germany}

\begin{abstract}

Here we evaluate the many-body entanglement properties of a generalized SU($n$) valence bond solid state on a chain. Our results follow from a derivation of the transfer matrix of the system which, in combination with symmetry properties, allows for a new, elegant and straightforward evaluation of different entanglement measures. In particular, the geometric entanglement per block, correlation length, von Neumann and R\'enyi entropies of a block, localizable entanglement and entanglement length are obtained in a very simple way. All our results are in agreement with previous derivations for the SU($2$) case. 

\end{abstract}


\maketitle

{\it Introduction.-} The study of entanglement in strongly correlated systems has proven fruitful to understand new phases of quantum matter and new types of quantum order \cite{wen}. In this respect, there has been growing interest in quantifying entanglement in the ground state of quantum many-body systems in one spatial dimension \cite{1d}. At criticality, the entanglement in these systems diverges, in turn obeying precise scaling laws orchestrated by the underlying conformal symmetry. Away from criticality, though, the existence of a finite correlation length and a non-zero gap to excitations forces entanglement to remain finite. 

The archetypical example of a quantum spin chain with a gap is the spin-1 AKLT model, introduced in Ref.\cite{aklt} by Affleck, Kennedy, Lieb and Tasaki. This model is invariant under rotations, that is, SU(2) operations. Moreover, its ground state is a valence bond solid (VBS) that admits a representation in terms of a Matrix Product State (MPS) \cite{mps}, and is closely related to the Laughlin state \cite{lau} and the fractional quantum Hall effect \cite{hall}. This scenario has been recently generalized to other symmetry groups such as SO($n$), SU($n$) and Sp($2n$) \cite{son,sun,sun1,sun2,sun3,sp2n}. As for the behaviour of entanglement in these generalizatons, not too much is known. Derivations have been carried out for the correlation length \cite{sun,sun1} as well as von Neumann and R\'enyi entropies \cite{sun2,sun3} of SU($n$) VBS states on a chain, but these involve a number of technicalities that make them quite lengthy. 

In this paper we provide an elegant and straightforward evaluation of the many-body entanglement properties of the above SU($n$) valence bond solid state on a chain. In particular, we derive unknown quantities such as the geometric entanglement per block \cite{ge,geb}, but also re-derive other quantities such as the correlation length, von Neumann and R\'enyi entropies of a block in a significantly simpler way than previous derivations \cite{sun2,sun3}. Our calculations are novel in many aspects for SU($n$) VBS states and are based on a proper understanding of (i) the structure of the MPS transfer matrix of a block, and (ii) the constraints imposed by SU($n$) symmetry. Furthermore, we also consider the localizable entanglement \cite{loca} and prove that the entanglement length diverges in the thermodynamic limit \cite{loca2}. 

\begin{figure}
\includegraphics[width=8.5cm,angle=0]{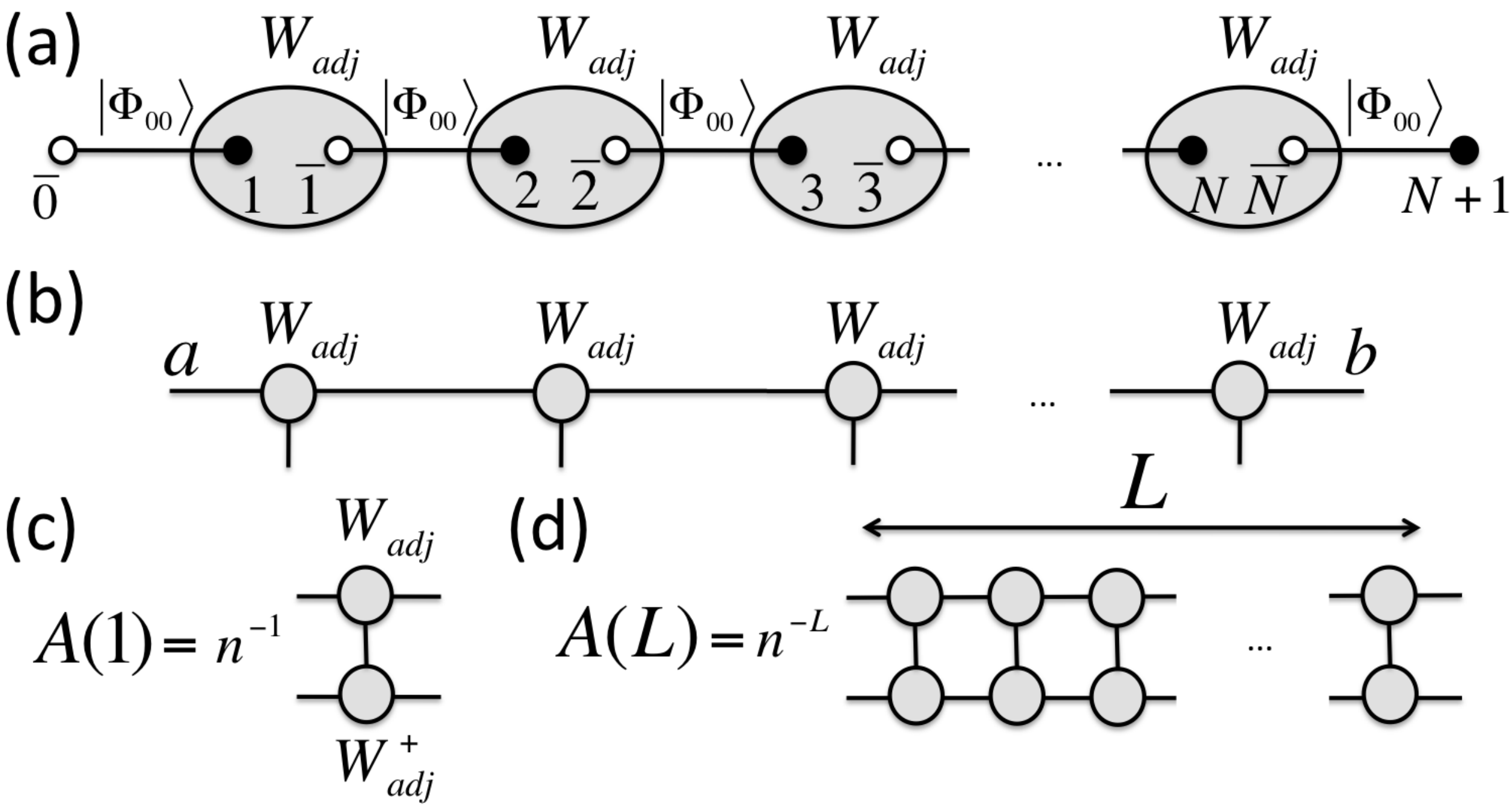}
\caption{(a) Structure of the SU($n$) VBS state $\ket{\Psi}$ for $N$ sites and open boundary conditions. (b) MPS structure of the state (up to an overall normalization constant). At the boundaries there are SU($n$) degrees of freedom $a,b = 0, 1, \ldots, n-1$. (c) MPS transfer matrix $A(1)$. (d) Transfer matrix of a block $A(L)=A(1)^L$.} 
\label{fig1}
\end{figure}

{\it The SU($n$) VBS state:} We start by reviewing the construction of the one-dimensional VBS state with SU($n$) symmetry from Refs.\cite{sun2,sun3}. This is the SU($n$) analogue of the ground state of the spin-1 AKLT chain, and can be represented for $N$ sites as
\beq
\ket{\Psi} = \left(\otimes_{r=1}^N W_{adj}^{[r,\bar{r}]} \right) \ket{\Phi_{00}}^{[\bar{0},1]}  \ket{\Phi_{00}}^{[\bar{1},2]} \cdots  \ket{\Phi_{00}}^{[\bar{N},N+1]} \ .
\eeq
In the above equation, $\ket{\Phi_{00}}^{[\bar{r},r+1]}$ is a singlet state of two `virtual' particles at sites $\bar{r}$ and $r+1$, one in the conjugate ($\widebar{\square}$)  and one in the fundamental ($\square$) representations of SU($n$), both of dimension $n$. Operator $W_{adj}^{[r,\bar{r}]}$ is the projector onto the $(n^2-1)$-dimensional adjoint subspace of $\square \otimes \widebar{\square}$. This adjoint subspace is the physical space at site $r$. As explained in Refs.\cite{sun1,sun2,sun3}, state $\ket{\Psi}$ is the ground state of a gapped parent Hamiltonian with SU($n$) symmetry. In fact, for $n>2$ state $\ket{\Psi}$ violates parity symmetry since the fundamental and conjugate representations of SU($n$) are different. This implies that for open boundary conditions it is the unique ground state of its parent Hamiltonian, whereas for periodic boundary conditions there are two degenerate ground states (one for each parity sector) depending on whether we choose the singlets $\ket{\Phi_{00}}$ for the tensor products $\widebar{\square} \otimes \square$ or $\square \otimes \widebar{\square}$. The VBS state $\ket{\Psi}$ is represented diagramatically in Fig.(\ref{fig1}.a). 

Let us elucidate further the structure of the above state. First, notice that by assigning a basis $\ket{j}$ to the fundamental representation and $\ket{\bar{j}}$ to the conjugate (with $j = 0, 1, \ldots, n-1$), the singlet $\ket{\Phi_{00}} $ can always be represented as a maximally entangled state, i.e. $\ket{\Phi_{00}} = \frac{1}{\sqrt{n}}\sum_{j=0}^{n-1}\ket{\bar{j}} \ket{j} $. Second, we remind that the tensor product of the fundamental and conjugate representations of SU($n$) obeys $\square \otimes \widebar{\square} = {\rm singlet} \oplus {\rm adjoint} $, where by `singlet' we mean the singlet representation (of dimension $1$) and by `adjoint' we mean the adjoint representation (of dimension $n^2-1$). In terms of a completeness relation for vector spaces, this implies that
\beq
\mathbb{I}_n \otimes \bar{\mathbb{I}}_n = \ket{\Phi_{00}}\bra{\Phi_{00}} + W_{adj} \ ,
\label{comp}
\eeq
where $\mathbb{I}_n = \sum_{j=0}^{n-1} \ket{j} \bra{j}$ is the $n \times n$ identity matrix between two fundamental spaces, and analogously for $\bar{\mathbb{I}}_n$. Thus, we can understand operator $W_{adj} = \mathbb{I}_n \otimes \bar{\mathbb{I}}_n - \ket{\Phi_{00}}\bra{\Phi_{00}}$ as a projector that removes the singlet component of the tensor product $\square \otimes \widebar{\square}$. 

At this point, we wish to introduce a very convenient notation in terms of tensor network diagrams: \emph{we will draw a line every time we wish to represent an identity matrix}. This identity matrix will be \emph{between two fundamental spaces, two conjugate spaces, or one fundamental space and one conjugate space}, with no distinction from the diagrammatic point of view. Using this notation, Eq.(\ref{comp}) reads 
\beq
~ \mbox{\boldmath \LARGE $)($} ~  = \frac{1}{n} ~ \mbox{\boldmath \LARGE $\asymp$} + W_{adj} \ .
\label{compdig}
\eeq
In the above expression, the left hand side is nothing but a diagram representing $\mathbb{I}_n \otimes \bar{\mathbb{I}}_n$. To clarify the meaning of the right hand side, notice that the singlet state must be represented as $\ket{\Phi_{00}} = [1/\sqrt{n}] ~ \mbox{\boldmath \Large $\frown$}$. We thus represent the projector in the singlet subspace as $\frac{1}{n} ~ \mbox{\boldmath \LARGE $\asymp$}$, which is the first term in the right hand side of Eq.(\ref{compdig}). With this notation it is also possible to represent the MPS structure of state $\ket{\Psi}$ in a clear and elegant way as shown in Fig.(\ref{fig1}.b), where the MPS matrices are just proportional to the projectors $W_{adj}$ at every site \footnote{Actually, there is also a representation where these matrices are the generators of the Lie algebra of SU($n$), $t^a$, $a = 1, \ldots, n^2-1$.}.  

Formally, we can think of Eq.(\ref{compdig}) as the completeness relation for the tensor product of `left' and `right' SU($n$) representations. From a geometrical perspective, though, one can also have the tensor product of `up' and `down' representations. In this case, one has the relation
\beq
~ \mbox{\boldmath \LARGE $\asymp$} ~  = \frac{1}{n} ~ \mbox{\boldmath \LARGE $)($} + W^\prime_{adj} \ ,
\label{compdig2}
\eeq
with $W^\prime_{adj}$ the projector onto the adjoint subspace of the considered tensor product. Both Eqs.(\ref{compdig}) and (\ref{compdig2}) will turn out to play a key role in our derivations. 

{\it Transfer matrix of a block:} All the calculations in this paper rely on the derivation of the MPS transfer matrix $A(1)$ for state $
\ket{\Psi}$. This is defined in terms of the MPS matrices $W_{adj}$ as $A(1) = (1/n)W_{adj} W_{adj}^{\dagger}$, see Fig.{(\ref{fig1}.c)}. Since $W_{adj}$ is a projector, we have that $W_{adj} W_{adj}^{\dagger} =  W_{adj} W_{adj} =  W_{adj}^2 = W_{adj}$, which can be checked easily using e.g. Eq.(\ref{compdig}). So, in the end, we obtain the remarkably simple expression
\beq
A(1) = \frac{1}{n} ~ \mbox{\boldmath \LARGE $)($} ~  - \frac{1}{n^2} ~ \mbox{\boldmath \LARGE $\asymp$} \ .
\label{tm}
\eeq

We now wish to compute the transfer matrix $A(L)$ of a block of length $L$. As shown in Fig.{(\ref{fig1}.d)} this is given by $A(L) = A(1)^L$, and thus can be obtained easily after diagonalizing $A(1)$ with respect to `left' and `right' spaces. This diagonalization is indeed straightforward if one realizes that the transfer matrix $A(1)$ in Eq.(\ref{tm}) can be rewritten using Eq.(\ref{compdig2}) as
\beq
A(1) = \left[1-\frac{1}{n^2}\right] \frac{1}{n} ~  \mbox{\boldmath \LARGE $)($} ~ + \left[\frac{-1}{n^2}\right] W^\prime_{adj} \ ,
\label{tmd}
\eeq
which is a sum of orthogonal projectors in singlet and adjoint subspaces, and is thus diagonal. Its largest eigenvalue is non-degerenate and is given by $\lambda_1 = (1-1/n^2)$ corresponding to the singlet eigenspace. There is also a second eigenvalue $\lambda_2 = -1/n^2$ with degeneracy $n^2 - 1$ corresponding to the adjoint eigenspace. Using this, the transfer matrix of a block reads
\beq
A(L) = \left[1-\frac{1}{n^2}\right]^L \frac{1}{n} ~  \mbox{\boldmath \LARGE $)($} ~ + \left[\frac{-1}{n^2}\right]^L W^\prime_{adj} \ .
\label{tmb}
\eeq

As expected, the above expression for $A(L)$ is manifestly SU($n$) invariant. To see this, just notice that thanks to Eq.(\ref{compdig2}) one can regard $A(L)$ as a combination of tensor products of $n \times n$ identity operators, which is of course invariant under any SU($n$) transformation.  This, in fact, is a consequence of the SU($n$) invariance of the quantum state $\ket{\Psi}$. 

{\it Correlation length and state norm.-} Some properties of the SU($n$) VBS state $\ket{\Psi}$ can already be computed just by looking at Eqs.(\ref{tmd}) and (\ref{tmb}). For instance, the correlation length is given by $\xi_C = -1/\log{|\lambda_2/\lambda_1|} = 1/\log{(n^2 - 1)} $, in agreement with results present in the literature  \cite{sun,sun1}. Furthermore, we can also compute the norm $\braket{\Psi}{\Psi}$ of the state, which will be needed in the forthcoming calculations. For a system of size $N$ and open boundary conditions this is given by $\braket{\Psi}{\Psi} = n\lambda_1^N$, whereas for periodic boundary conditions $\braket{\Psi}{\Psi} = \lambda_1^N + (n^2-1) \lambda_2^N$, with $\lambda_1$ and $\lambda_2$ defined as above. In the thermodynamic limit the norm scales as $N^{-1} \log{\braket{\Psi}{\Psi}} \sim \log{\lambda_1} = \log{\left(1-1/n^2\right)}$ regardless of the boundary condition. 

{\it Geometric entanglement per block.-} We now consider the behaviour of the geometric entanglement per block of length $L$, which we call $\mathcal{E}(L)$ \cite{geb}. This is defined in the thermodynamic limit as $\mathcal{E}(L) \equiv \lim_{(N/L) \rightarrow \infty} (N/L)^{-1} E(\Psi)$, with $E(\Psi) = - \log{(|\Lambda_{{\rm max}}|^2)}$ and $|\Lambda_{{\rm max}}|^2 = {\rm max}\left| \braket{\Phi}{\Psi}/\sqrt{\braket{\Psi}{\Psi}}\right|^2 $, where the maximization is done over all possible product states  $\ket{\Phi}$ of contiguous blocks. This measure of entanglement quantifies the resemblance of the quantum state $\ket{\Psi}$ to a product state of the blocks. Thus, studying how $\mathcal{E}(L)$ scales with $L$ allows us to understand how similar $\ket{\Psi}$ is to a product state under successive coarse-grainings of the system \cite{geb}. In our case, since the system is gapped, we expect $\mathcal{E}(L)$ to saturate in some constant for $L \gg \xi_C$. This will turn out to be the case, as we shall see shortly. 

Let us now perform the calculation of $\mathcal{E}(L)$. Our derivation follows similar lines as the calculations in Ref.\cite{geaklt}. First of all, since the system is antiferromagnetic the basic unit cell is $2$, and thus we consider here the case of blocks of \emph{even} length $L$ only \footnote{The case of odd $L$ needs of some alternative derivation and is not considered in this paper.}. Second, as explained in Refs.\cite{geaklt} and \cite{gexi}, for an MPS with diagonalizable transfer matrix $A(1)$ the quantity $|\Lambda_{{\rm max}}|^2$ is given in the limit $N \gg 1$ by $|\Lambda_{{\rm max}}|^2 \sim \left({\rm max} |(r|^* (r| A(L) |r) |r)^* | \right)^{N/L} / \braket{\Psi}{\Psi} $, where $|r), |r)^*, (r|^*$ and $(r|$ are normalized vectors in the four spaces connected by $A(L)$ (respectively right-up, right-down, left-up and left-down), and the maximization is done over these vectors. Generally speaking, carrying this maximization is cumbersome. However, an important simplification occurs in our case since $A(L)$ is just a combination of tensor products of identity operators. This implies that the maximization \emph{does not depend on vector $|r)$ at all}, which is a consequence of the SU($n$) symmetry of $A(L)$ \footnote{This was also observed for the SU(2) case in Ref.\cite{geaklt}.}. Using Eq.(\ref{tmb}) and our previous result for the norm, one can then see that
\beq
|\Lambda_{{\rm max}}|^2 \sim \left(\frac{(1-1/n^2)^L(1/n) + (-1/n^2)^L (1-1/n)}{(1-1/n^2)^L} \right)^{N/L} 
\eeq
which means that the geometric entanglement per block is ultimately given by 
\beq
\mathcal{E}(L) = \log{n} - \log{\left(1 + (n-1)e^{-L/\xi_C}\right)} ~~~ L ~ {\rm even} \ , 
\label{ge2}
\eeq
where we have used our previous result for the correlation length $\xi_C$. As expected, the above formula matches the SU(2) case calculated in Ref.\cite{geaklt} for $n=2$. Also, for $L \gg \xi_C$ the geometric entanglement saturates in $\mathcal{E}(L \gg 1) \sim \log{n}$, as expected for a gapped system. We also see that for $n \gg 1$ the geometric entanglement increases linearly with the size of the block, $\mathcal{E}(L) \sim L/\xi_C$, before reaching the saturation regime for large $L$. 

{\it Von Neumann and R\'enyi entropies of a block.-} Our derivation of the transfer matrix of a block allows us to compute as well the Von Neumann and R\'enyi entropies of a block, $S(L)$ and $S_{\alpha}(L)$. These are defined respectively as $S(L) \equiv -\sum_a\lambda^\prime_a \log \lambda^\prime_a $ and $S_{\alpha}(L) \equiv (1-\alpha)^{-1}\log{\left( \sum_a(\lambda^\prime_a)^{\alpha} \right)}$ for $\alpha \neq 1$ and $\alpha > 0$, and where $\lambda^\prime_a$ are the eigenvalues of the reduced density matrix of a block of size $L$, which we call $\rho_L$. The set of R\'enyi entropies contains lots of information about the entanglement in the system. Both the von Neumann and R\'enyi entropies were already computed in Ref.\cite{sun2} for the SU($n$) VBS state $\ket{\Psi}$ using a different method. Here, though, we re-derive them by using the formalism introduced earlier, which allows for a new, simple and elegant derivation in just a few lines. 

\begin{figure}
\includegraphics[width=8cm,angle=0]{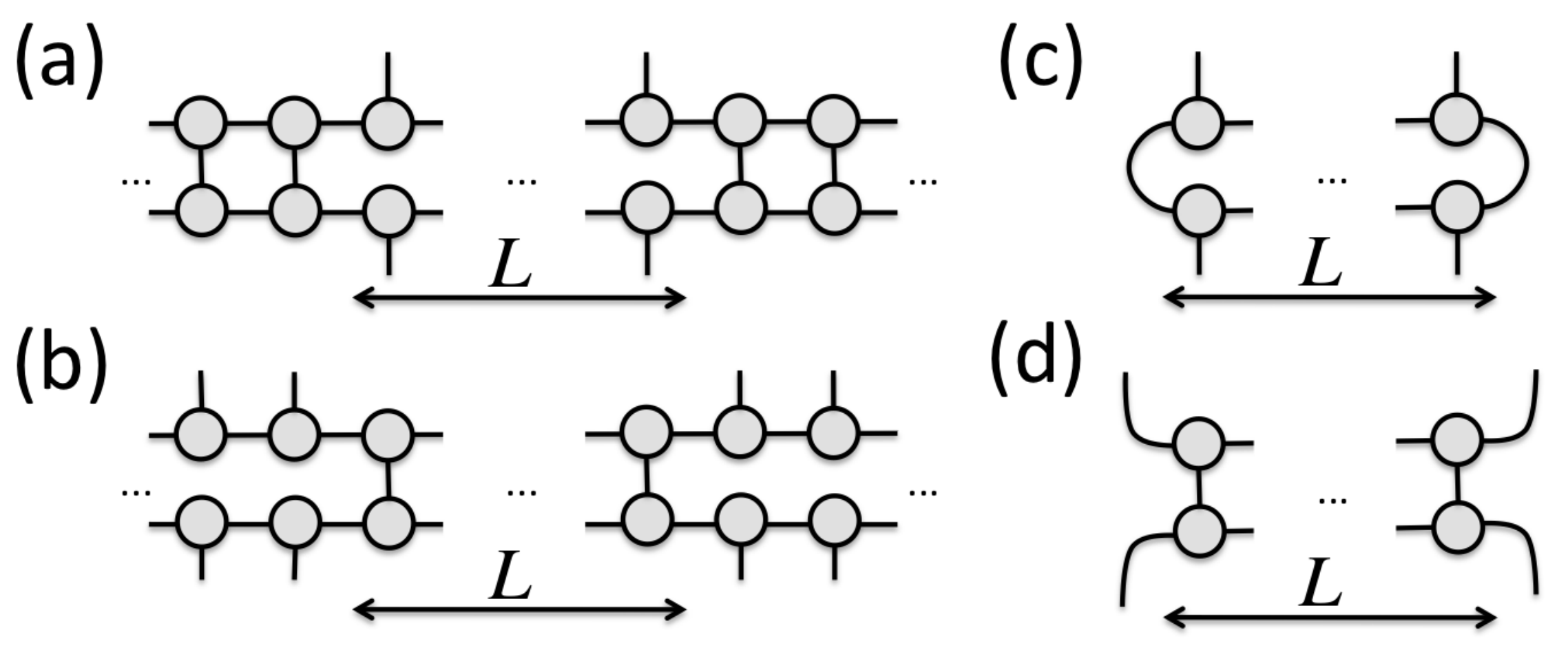}
\caption{Diagrammatic representation of several reduced density matrices sharing the same spectrum (up to normalization constants). (a) Of a block of length $L$, called $\rho_L$. (b) Of the environment of a block of length $L$ for the environment as in (a). (c) Of a block of length $L$ once the environment is computed (dominant left and right eigenvectors of $A(1)$, which are singlets). (d) Of the environment of a block of length $L$ for the environment as in (c), equal to $A(L)$.} 
\label{fig2}
\end{figure}

The calculation of these entropies amounts to being able to diagonalize the reduced density matrix of a block $\rho_L$. In terms of tensor network diagrams, this reduced density matrix is represented in Fig.(\ref{fig2}.a,c) (up to normalization). The key observation is that the eigenvalues of $\rho_L$ must be equal to those of the reduced density matrix of the \emph{environment} of the block, see Fig.(\ref{fig2}.b,d). Thus, up to a normalization constant, we can write
\beq
\rho_L \leftrightarrow A(L) \ ,
\eeq
where $\leftrightarrow$ means that they share the same eigenvalues if we diagonalize $A(L)$ with respect to the `up' and `down' spaces. 
This diagonalization can be done by using again the decomposition given in Eq.(\ref{compdig}). We obtain the diagonal expression
\beqa
A(L) &=&  \left[ \left(1 - \frac{1}{n^2}\right)^L \frac{1}{n} + \left(\frac{n^2-1}{n}\right)\left(\frac{-1}{n^2}\right)^L\right] \frac{1}{n} \mbox{\boldmath \LARGE $\asymp$} \nonumber \\
&+& \left[ \left(1 - \frac{1}{n^2}\right)^L \frac{1}{n} - \left(\frac{1}{n}\right)\left(\frac{-1}{n^2}\right)^L\right] W_{adj}  \ .
\eeqa
Defining $p_n(L) \equiv (-1/(n^2-1))^L = (-1)^L \exp{(-L/\xi_C)}$, we find the normalized eigenvalues
\beqa
\lambda^\prime_1 &=& \frac{1}{n^2}(1 + (n^2-1)p_n(L))  \nonumber \\
\lambda^\prime_2 &=& \frac{1}{n^2}(1 - p_n(L)) \ ,
\label{ei}
\eeqa
where $\lambda^\prime_1$ has degeneracy 1 (singlet eigenspace) and $\lambda^\prime_2$ has degeneracy $n^2-1$ (adjoint eigenspace). These eigenvalues coincide exactly with the ones given in Eq.(24) of Ref.\cite{sun2}, and from them the von Neumann and R\'enyi entropies follow inmediately. Notice that this derivation is also far more straightforward than the one in Ref.\cite{sun2} since, essentially, we just used repeated applications of Eq.(\ref{comp}) in different contexts. As expected, for the $n=2$ case one recovers the expressions for the spin-1 AKLT state in Ref.\cite{akko}. Furthermore, in the limit $L\gg \xi_C$ we see that $\lambda^\prime_1 \sim \lambda^\prime_2 \sim 1/n^2$. In this limit the reduced density matrix $\rho_L$ is thus proportional to some projector on an $n^2$-dimensional subspace, and the von Neumann and R\'enyi entropies saturate as $S(L) \sim S_{\alpha}(L) \sim 2 \log{n}$, as was also observed in Ref.\cite{sun2}. Interestingly, this saturation value is twice the one that we obtained for the geometric entanglement per block in the same limit. 

{\it Localizable entanglement and entanglement length.-} As in the SU($2$) case \cite{loca2}, it is also possible to compute exactly the localizable entanglement and entanglement length of this system. By definition, the localizable entanglement is the maximum entanglement that can be localized, on average, between two parties of a multipartite system, by doing local measurements on the other parties. This brings to the concept of entanglement length $\xi_E$ as the length scale at which it is possible to create a maximally entangled state between two parties by doing measurements on the rest of the parties. In our case the parties are the different sites of the system, corresponding to adjoint representations of SU($n$). 

Our derivation of $\xi_E$ is a generalization of the one in Ref.\cite{loca2}. Specifically, we consider the generalized Bell basis for two qu-nits $\ket{\Phi_{lp}} = \frac{1}{\sqrt{n}} \sum_{j=0}^{n-1} e^{2 \pi i l j/n} \ket{j \oplus p} \ket{\bar{j}} $, with $l,p = 0, 1, \ldots, n-1$. Notice that for $l=0, p=0$ we recover the singlet state $\ket{\Phi_{00}}$. A measurement in this basis achieves perfect teleportation for the state of one qu-nit \cite{tele}. The key observation is that, of the above states, $\ket{\Phi_{00}}$ is the singlet, whereas the rest of states are an orthonormal basis of the adjoint $(n^2-1)$-dimensional subspace. Thus, \emph{a measurement at each site on this basis for $l,p \neq 0$ amounts to a generalized Bell measurement on the virtual particles}. As in Ref.\cite{loca2} these measurements implement entanglement swapping, which means that the localizable entanglement between any pair of sites is always maximum. In particular, such measurements allow to create a maximally entangled state between the two qu-nits at the boundaries of the system. Therefore, for $N$ sites we have that $\xi_E \sim N$, which diverges in the thermodynamic limit. Importantly, and as in Ref.\cite{loca2}, the entanglement length $\xi_E$ diverges while the correlation length $\xi_C$ is finite. 

{\it Conclusions and outlook.-} Here we have shown how to obtain the entanglement of a VBS state on a chain with SU($n$) symmetry in a novel, elegant and straightforward way, much simpler than previous derivations. We believe that similar techniques to the ones in this paper could also be useful in the study of one-dimensional VBS states with other symmetries such as SO($n$) and Sp($2n$) \cite{son,sp2n}, as well as SU($n$) VBS states in higher dimensions \cite{sunhigh}. 

We acknowledge discussions with I. Cirac and H. Katsura. R.O. acknowledges support from the EU through a Marie Curie International Incoming Fellowship.

\end{document}